\documentclass[showpacs,aps,prl,twocolumn]{revtex4}
\usepackage{graphicx}
\usepackage{dcolumn}
\usepackage{hyperref}
\usepackage{psfrag,bm,amsmath,amsfonts,amssymb,color}
\usepackage{times}
\usepackage{mathrsfs}
\begin{document}

\def\sss{\scriptscriptstyle}

\title{Direct measurement of evolving dark energy density and super-accelerating expansion of the universe}

\author{Shuang-Nan Zhang$^{1,2}$}\email{zhangsn@ihep.ac.cn}
\author{Yin-Zhe Ma$^{3,4}$}

\affiliation{
$^1$Key Laboratory for Particle Astrophysics, Institute of High Energy Physics, CAS, 19B Yuquan Road,
Beijing 100049, China}
\affiliation{
$^2$National Astronomical Observatories of China, CAS, 20A Datun Road, Beijing 100020, China}
\affiliation{
$^3$Department of Physics and Astronomy, University of British Columbia, Vancouver, BC, Canada}
\affiliation{
$^4$Canadian Institute for Theoretical Astrophysics, Toronto, Ontario, Canada}

\date{\today}

\begin{abstract}
A higher value of Hubble constant has been obtained from measurements with nearby Type Ia supernovae, than that obtained at much higher redshift. With the peculiar motions of their hosts, we find that the matter content at such low redshift is only about 10\% of that at much higher redshifts; such a low matter density cannot be produced from density perturbations in the background of the $\Lambda$CDM expansion. Recently the \textit{Planck} team has reported a lower Hubble constant and a higher matter content. We find that the dark energy density increases with cosmic time, so that its equation-of-state parameter decreases with cosmic time and is less than -1 at low redshift. Such dark energy evolution is responsible for driving the super-accelerating expansion of the universe. In this extended $\Lambda$CDM model, the cosmological redshift represents time rather than radial coordinate, so that the universe complies to the Copernican Principle.
\end{abstract}
\pacs{98.80.Es}

\maketitle

The Hubble constant $H_0$ measures the expansion rate of present day universe, provides the basic information on the age of the universe, and is a key
parameter related to other cosmological parameters, such as densities of dark matter (DM) and dark energy (DE) in the universe. $H_0$ can be determined by measuring the Hubble parameter $H(z)\equiv \dot{a(z)}/a(z)$ at any redshift $z$ and then projecting it to $z=0$ with an underlying cosmological model, where $a$ is the scale factor. Therefore $H_0$ determined this way is model-dependent, unless $z\approx 0$. In the following, $H_{0,z}$ denotes $H_0$ projected with measurements at $z$. This means, in principle, only $H_{0,0}$ is model-independent. The best model-independent measurement of $H_{0,0}$ can be made using nearby Type Ia supernovae (SNe~Ia). Recently a 3.3\% error of $h_{0,0}=0.738$ ($h\equiv H/100$~km~s$^{-1}$~Mpc$^{-1}$) is reported by calibrating these standard candles with many Cepheids in
their host galaxies \cite{Riess2011,Riess2012}.

Currently the standard cosmology model is the base $\Lambda$CDM model, in which the cold DM and DE ($\Lambda$) dominate the matter and energy contents of the universe and the DE density ($\rho_{\Lambda}$) does not change with cosmic time ($t$). Decisive evidence for the existence of DE was found from comparisons between the apparent magnitudes of the low- and high-$z$ SNe~Ia, which led the discovery of the accelerating expansion of the universe \cite{Riess1998,Perlmutter1999}. The projected $H_0$ with the base $\Lambda$CDM model should be the same from measurements made at all $z$, if the base $\Lambda$CDM model is valid at all $z$ with universal parameters. For example, $\rho_{\Lambda}\equiv 3H_0^2\Omega_\Lambda/8\pi G$  should remain the same from measurements made at all $z$, since $H_0$ and $\Omega_\Lambda$ are universal in the base $\Lambda$CDM model. For convenience we define the normalized DE parameter $\Psi_{\Lambda,z}=\Omega_{\Lambda,z} h_{0,z}^2$, where $\Omega_{\Lambda,z}$ and $h_{0,z}$ are obtained with measurements at $z$; we then have $\rho_{\Lambda,z}=\frac{3\times \tilde{H}_0^2 }{8\pi G}\Psi_{\Lambda,z}$, where $\tilde{H}_0=100$~km~s$^{-1}$~Mpc$^{-1}$. Since $\rho_{\Lambda,z}$ does not vary in the base $\Lambda$CDM model, $\rho_{\Lambda,z}$ obtained by fitting data with the base $\Lambda$CDM model is actually $\rho_{\Lambda}$ at $z$. Therefore the base $\Lambda$CDM model provides a convenient framework to determine directly the evolution of $\rho_{\Lambda}$. Since observationally we normally measure $\Omega_{{\rm M},z}h_{0,z}^2$ at $z$, we re-write $\Psi_{\Lambda,z}=(1-\Omega_{{\rm M},z})h_{0,z}^2$, for a flat universe ($\Omega_{K}=0$). Recently, the {\textit{Planck}} team has released their results as $h_{0,z}=0.679\pm 0.015$ and $\Omega_{{\rm M},z}h^2=0.1423\pm0.0029$ (at $z\sim 1100$) \cite{PlanckCollaboration2013}. As we will show in this paper, both of them are statistically different from the low-$z$ measurements of SNe~Ia, requiring an increasing $\rho_{\Lambda}$ with $t$.

\begin{figure}
\begin{center}
\includegraphics[angle=0,width=0.4\textwidth]{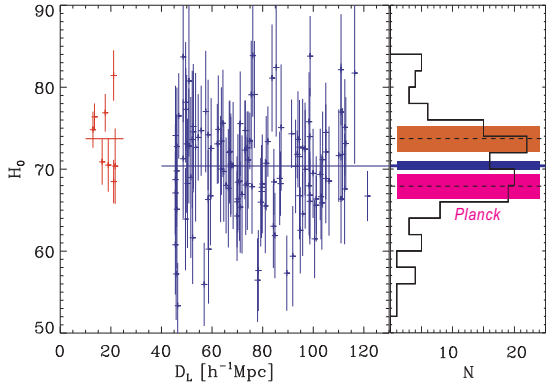}
\end{center}
\caption{$h_0$ measured with each SNe~Ia within a luminosity distance $D_L$ of ~120~$h^{-1}$Mpc.
{\bf Left panel:} The red crosses are the eight SNe~Ia \cite{Riess2011} used to measure the local $H_0$ with
$D_L<25$~$h^{-1}$Mpc, giving an average local $h_{0,0}=0.738\pm 0.0155$ marked as the thick solid red line. The blue crosses are the Union 2.1 SNe~Ia \cite{Suzuki2012} at $D_L>40$~$h^{-1}$Mpc, yielding an average $h_{0,z_{1-}}=0.704\pm 0.0051$ as the thick solid blue line ($z_{1-}=0.025$ is the median $z$ of these SNe~Ia marked by the blue crosses). Seven of the eight SNe~Ia (red crosses) have $h_0>h_{0,z_{1-}}$, indicating that the probability that the eight SNe~Ia are drawn from the same population of the other SNe~Ia (blue crosses) is less than 3.6\%. The null hypothesis that the two samples have the same mean is rejected at 96.3\% confidence level with Welch's $t$-test. {\bf Right panel:} Histogram of the blue crosses in the left panel. The filled red and blue areas are the $1\sigma$ error regions of $h_{0,0}$ and $h_{0,z_{1-}}$ respectively; their errors are calculated from the variance of each sample, and are significantly larger than that calculated from error propagation using the measurement errors of all data points (see text for details). The large error in $h_{0,0}$ is due to its small sample size and additional fluctuations caused by the peculiar motions of their hosts (see text for details). $h_{0,0}$ and $h_{0,z_{1-}}$ are different at $2.1\sigma$ level with respect to their joint error bar, i.e., the probability that they are consistent with each other is less than 3.6\%. For comparison, the just released {\textit{Planck}} result $h_{0, z_2}=0.679\pm 0.015$ is also marked by the filled magenta area ($z_2\sim 1100$).}
 \label{fig:1}
\end{figure}

Indeed, just before the discovery of the accelerating expansion
of the universe \cite{Riess1998,Perlmutter1999}, evidence was found that $H_{0,0}>H_{0,z}$ ($z \gtrsim 0.01$) by about 6\%, where the boundary is around $D_L\sim70$~$h^{-1}$Mpc \cite{Zehavi1998}. This suggests that we are living within a local bubble expanding slightly faster than the outside universe. Therefore we are moving away with respect to distant SNe~Ia faster than the global Hubble expansion and thus distant SNe~Ia should look dimmer than viewed only within the global Hubble flow. This has led heated debates if the accelerating expansion of the universe is simply an mirage of this local bubble \cite{Alnes2006,Alexander2009,Mattsson2009,Mortonson2009,Sinclair2010,Moss2011,Ellis2011,Nadathur2011,Marra2011}, since the over-dimming of distant SNe~Ia is what led to the initial discovery of the accelerating expansion of the universe. However in the base $\Lambda$CDM model ($\rho_{\Lambda,z}=const$), $H_{0,0}$ should be considered a global property of the universe, and can be used directly as a pre-determined parameter when constraining the other cosmological parameters with cosmic microwave background (CMB) observations ($z\sim 1100$) \cite{Hinshaw2012}. As we will show in this paper, neither model agrees with the SNe~Ia data with $z \gtrsim 0.01$ and thus the void model is rejected and the base $\Lambda$CDM model should be extended by allowing evolving $\rho_{\Lambda}$ with $t$.

In Figure~\ref{fig:1} we show $H_0$ measured with the same eight SNe~Ia ($z=0.0043$ to $0.0072$) used to obtain $H_{0,0}$ \cite{Riess2011} and the Union 2.1 compilation \cite{Suzuki2012} to measure $H_{0,z}$. The data for the eight SNe~Ia are listed in Table~1 in the Supplementary Material (SM); $h_{0}$ (with standard error $\sigma_{h_{0}}$) is calculated using Equation~(4) and data in Table~3 of Ref.\cite{Riess2011}. $h_{0}$ for other SNe~Ia is calculated using Equation (12) or (13) in Ref.\cite{Riess2009} with the same parameters. We limit the SNe~Ia with $z<0.04$ (with a median $z_{1-}=0.025$), in order to avoid any coupling with cosmological parameters \cite{Riess2009}. We obtain $h_{0,0}=0.738\pm 0.0155$ and $h_{0,z_{1-}}=0.704\pm 0.0051$: $h_{0,0}>h_{0,z_{1-}}$ at 96.4\% confidence level (CL). Since our goal here is to examine the statistical consistency between the two values of $h_0$, only statistical errors in these SNe~Ia are included here; the effects of possibly larger errors in $h_{0,0}$, including cosmic variance, are discussed later. This confirm the previous conclusion \cite{Zehavi1998} with updated data. For comparison, we also show $h_{0,z_2}=0.679\pm 0.015$ ($z_2\sim 1100$) in Figure~\ref{fig:1}, reported by the {\textit{Planck}} team \cite{PlanckCollaboration2013}.
\begin{figure}
\begin{center}
\includegraphics[angle=0,width=0.4\textwidth]{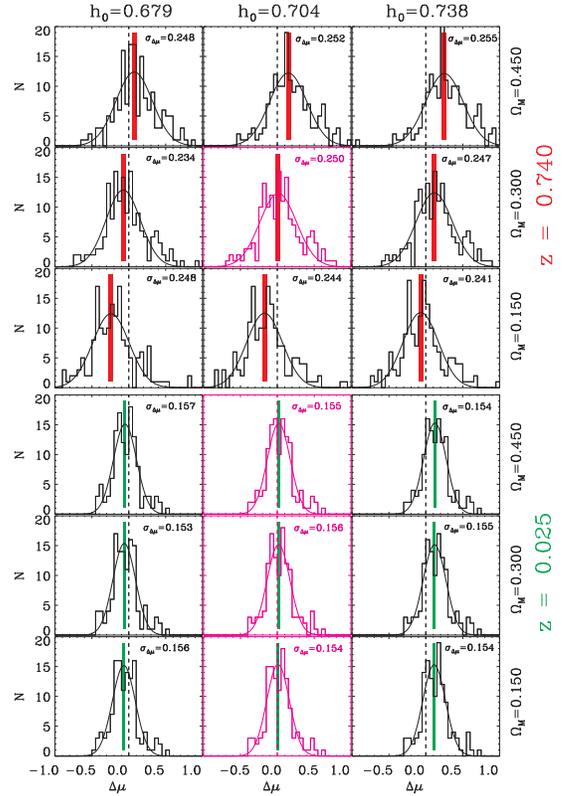}
\end{center}
\caption{Residuals of the distance modules ($\mu$) of the Union 2.1 SNe~Ia against the prediction of the $\Lambda$CDM
cosmological model with different combinations of $\Omega_{\rm M}$ and $H_0$; $\Omega_{\rm M}+\Omega_{\Lambda}=1$ is always assumed. The black thick vertical dashed lines indicate $\Delta\mu=0$. Two groups of SNe~Ia are chosen here: low-$z$ ($z\leq0.04$, with median of 0.025), and high-$z$ ($z\geq0.5$, with median of 0.740). The combination of the absolute value of $\widehat{\Delta\mu}$ (the center of the distribution of $\Delta\mu$) and the magnitude of $\sigma_{\Delta\mu}$ (the standard deviation of each Gaussian fit, also labelled in each panel), indicates how well the model describes the data. Each filled area marks the $3\sigma$ error range of $\widehat{\Delta\mu}$ with $\sigma_{\widehat{\Delta\mu}}=\sigma_{\Delta\mu}/\sqrt{n-1}$, where $n$ is the total number of data points. The four panels plotted in magenta color are consistent with data: $h_{0}=0.704$ for $z_{1-}=0.025$, independent of $\Omega_{\rm M}$; also $h_{0}=0.704$ for $z_{1+}=0.740$, unless $\Omega_{\rm M}$ deviates significantly from 0.3.}  \label{fig:4}
\end{figure}

\begin{figure}
\begin{center}
\includegraphics[angle=0,width=0.3\textwidth]{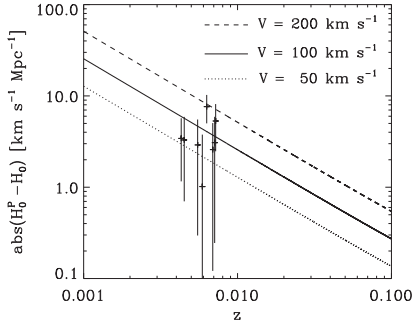}
\end{center}
\caption{Deviations of the measured $H_0$ with individual SNe~Ia, caused by the SNe~Ia peculiar motions. The lines are the expected deviations for different $V_{\rm los}$.
The data points with error bars are the deviations of the eight SNe~Ia from $h_{0,0}=0.738$. The data are consistent with $V_{\rm los}\sim 100$~km~s$^{-1}$.} \label{fig:2}
\end{figure}

In Figure~\ref{fig:4}, we examine critically if the three values of $h_0$, i.e., $h_{0,0}$, $h_{0,z_{1-}}$ and $h_{0,z_2}$, are consistent with the Union 2.1 data at both low- and high-$z$, within the framework of the base $\Lambda$CDM model with different values of (but not evolving) $\Omega_{\rm M}$. We conclude that only $h_{0}=0.704$ is consistent with data at low-$z$ ($z_{1-}=0.025$), independent of $\Omega_{\rm M}$. At high-$z$ ($z_{1+}=0.740$), again only $h_{0}=0.704$ is consistent with data unless $\Omega_{\rm M}$ deviates significantly from around 0.3; actually the high-$z$ SNe~Ia data favors a lower value of $\Omega_{\rm M}=0.28$. It is thus very unlikely that SNe~Ia data can be reconciled with either the higher or lower $h_0$ measured in the local bubble or with CMB data of {\textit{Planck}}. Therefore the combined SNe-Ia and {\textit{Planck}} data support an increasing $h_{0,z}$ with increasing $t$ or decreasing $z$. A remaining issue is whether the higher $h_{0,0}$ is just due to a density perturbation in the local universe, i.e., we are living in a local density void embedded in an otherwise unform expansion of the universe described by the base $\Lambda$CDM model. In such a scenario, the under-density is given by $-\Delta \Omega_{\rm M}/\Omega_{\rm M}=2\Delta h_0/h_0\approx 0.1$.

$\Omega_{\rm M}$ in the bubble can be measured by the peculiar velocity dispersion (PVD) of these SNe~Ia hosts. From each listed $H_{0,0}$ and its error in Table~1 in SM, the pure statistical error of $h_{0,0}$ should be 0.009, smaller than 0.0155 determined from the variance of the sample. This means that the
probability that the data do not contain additional fluctuations is less than 0.68\%. It has been known that SNe~Ia peculiar motions may cause such fluctuations
\cite{Hui2006,Zhang2008}. Since $H_0\cong {cz}/{D_L}$ when $z\cong 0$, a non-negligible deviation to $H_0$ may be produced for a peculiar velocity along the line of sight (LOS) $V_{\rm los}\sim 100$~km~s$^{-1}$ at $z\ll 1$. The additional fluctuations caused by random SNe~Ia peculiar motions can be found from $\sigma_{h_{\rm 0, P}}^2=\langle h_0^2\rangle-\bar
\sigma_{h_{0}}^2=0.0352^2$, where $\langle h_0^2\rangle=\sum (h_{0,i}-\bar h_0)^2/(n-1)$ and $\bar \sigma_{h_{0}}=\sum
\sigma_{h_{0,i}}/n=0.0262$ ($n=8$); in fact $\bar \sigma_{h_{0}}\simeq \sigma_{h_{0,i}}$. Clearly
$\sigma_{h_{\rm 0, P}}>\bar \sigma_{h_{0}}$, i.e., the average fluctuation to $h_0$ caused by
random SNe~Ia peculiar motions is larger than the measurement errors in $h_0$.

We then compare the measured $|H_{0,i}-\bar H_0|$ with the expected deviations caused by different $V_{\rm los}$ at low-$z$ in Figure~\ref{fig:2} ($\bar H_0=H_{0,0}$); the data are consistent with $V_{\rm los}\sim100$~km~s$^{-1}$. In Figure~1 of SM, we show the SNe~Ia peculiar motions do not show any significantly coordinated pattern (albeit with small number statistics) and thus are consistent with random motions \cite{Zehavi1998}. For random peculiar motions, the PVD between the eight SNe~Ia is found to be $\hat{\sigma}_{12}=141\pm26.5{\rm \ km \ s}^{-1}$ ($1\sigma$ range), with projected separations of these pairs between 5 to 40 $h^{-1}$Mpc (Please refer to SM). It has been shown that $\hat{\sigma}_{12}$ converges to $500{\rm \ km \ s}^{-1}$ (for $\Omega_{{\rm M}}=0.3$, $\hat{\sigma}_{12}\propto \Omega_{\rm M}^{0.55}$) at these separations \cite{Tinker2007}. We thus obtain $\Omega_{{\rm M},0}=0.030\pm 0.011$ and $-\Delta \Omega_{\rm M}/\Omega_{\rm M}\sim 0.9\gg 2\Delta h_0/h_0\approx 0.1$. This discrepancy cannot be reconciled even if the possible cosmic variance effect is considered on the uncertainty of $h_{0}=0.738$ with an additional $\sim2.5$\% \cite{Marra}. Therefore the void model is excluded with high significance. Actually the spherically symmetric but inhomogeneous Lema\^{\i}tre-Tolman-Bondi (LTB) model of the universe has been excluded with stringent limits \cite{Moss2011,Zhang2011c,Zibin2011,Wang2013a}.

\begin{figure}
\begin{center}
\includegraphics[angle=0,width=0.45\textwidth]{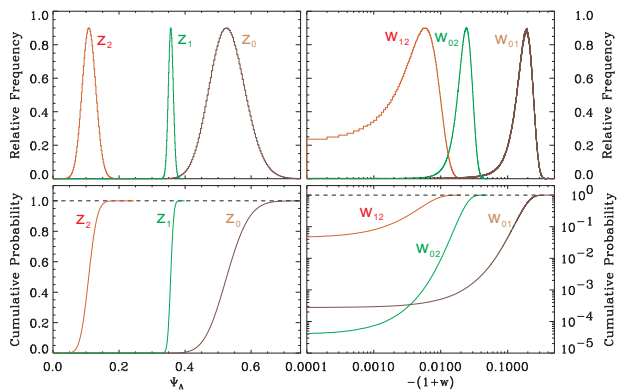}
\end{center}
\caption{The normalized DE density $\Psi_{\Lambda}$ and DE equation-of-state parameter $w$ obtained from measurements at three redshifts: $z_0\sim 0$, $z_1\sim 1$, and $z_2\sim 1100$. Input parameters with $1\sigma$-errors are: $h_{0,0}=0.703\pm0.0352$ (i.e., 5\% error in $h_{0,0}$), $\Omega_{{\rm M},0}=0.030 \pm 0.011$, $h_{0,z_1}=0.704\pm 0.0051$, $\Omega_{{\rm M},z_1}=0.28 \pm 0.01$, $h_{0,z_2}=0.679\pm 0.015$ and $\Omega_{{\rm M},z_2}h_{0,z_2}^2=0.1423\pm0.0029$. The cases for smaller errors in $h_0$ are discussed in SM; just one percent improvement in the precision to $h_0$ can improve the accuracy of $w$ considerably.} \label{psi_eos}
\end{figure}

As a straightforward and simple extension to the base $\Lambda$CDM model, the local bubble with $h_{0}=0.738$ and $\Omega_{\rm M}\sim0.03$ can be considered as the global property of present day universe; observationally it becomes ``local"  because only a small volume of present day universe can be observed by any observer, due to the finite light propagation speed. In other words, an observer located anywhere in the universe at $z\sim0$ should also observe the same local bubble. More specifically, the SNe~Ia and {\textit{Planck}} data support a scenario that the expansion rate of the present day universe is higher than that predicted in the base $\Lambda$CDM model, i.e., the projected $h_{0,z}$ increases with $t$. We call this {\it super-accelerating expansion of the universe}, to distinguish it from the well-known accelerating expansion of the universe described by the base $\Lambda$CDM model, with constant $\rho_{\Lambda}$ and $H_0$ \cite{Riess1998,Perlmutter1999}. The normalized DE densities at the three redshifts can be obtained directly as: $\Psi_{\Lambda,0}=0.53_{-0.10}^{+0.21}$ ($z_0\sim 0$),
 $\Psi_{\Lambda,z_1}=0.357_{-0.014}^{+0.029}$  ($z_1\sim 1$), and $\Psi_{\Lambda,z_2}=0.103_{-0.040}^{+0.082}$ ($z_2\sim 1100$); all errors quoted here are for 95\% CL (Figure~\ref{psi_eos}).

Matter and DE are completely decoupled in the $\Lambda$CDM model, therefore the energy conservation for DE requires $\rho_\Lambda\propto a^{-3(1+w)}$, where $w$ is the DE equation-of-state parameter. The cosmological model is commonly referred to as $w$CDM model, if $w$ is allowed to deviate from $-1$, as an extension to the base $\Lambda$CDM model. With measurements of $\Psi_{\Lambda}$ made at any two redshifts $z_i$ and $z_j$ ($z_i<z_j$), we have $3(1+w_{z_i,z_j})=-\log(\Psi_{\Lambda,z_i}/\Psi_{\Lambda,z_j})/\log((1+z_j)/(1+z_i))$. We therefore obtain: $-(w_{z_0,z_1}+1)=0.188_{-0.102}^{+0.194}$, $-(w_{z_0,z_2}+1)=0.024_{-0.011}^{+0.022}$, and
$-(w_{z_1,z_2}+1)=0.0060_{-0.0068}^{+0.014}$, respectively (Figure~\ref{psi_eos}). These results indicate that $P(w_{z_0,z_1}\ge -1)=1.5\times 10^{-4} (3.6\sigma)$, $P(w_{z_1,z_2}\ge -1)=4\times 10^{-2} (1.7\sigma)$ and $P(w_{z_0,z_2}\ge -1)=10^{-5} (4.3\sigma)$. Consequently the observed increasing $\rho_{\Lambda}$ with $t$ found here provides strong evidence for $w<-1$ (the so-called phantom DE) at low-$z$ and $w$ decreases with $t$. In this case the cosmological redshift $z$ can be interpreted as the time coordinate $t$ rather than radial coordinate $r$, and thus the universe can still maintain homogeneous and comply with the Copernican Principle (see SM for more details).

$\Omega_{\rm M}\sim0.03$ measured at $z= 0.0043$ to 0.0072, due to the low PVD of the SNe Ia hosts, seems to contradict the measured over-density just outside our local group \cite{Schlegel1994}. However, previous surveys of nearby galaxies have found that baryon density ($\Omega_{\rm b}$) declines very rapidly beyond the local group and is half of the cosmological average at $\sim$6 $h^{-1}$Mpc \cite{Karachentsev2004}, still inside the local supercluster. It is plausible that $\Omega_{\rm b}$ outside the local supercluster continues to decline by a factor a few, in agreement with very low $\Omega_{\rm M}$ in the local bubble, if baryon matter traces DM halos in present day universe. This may explain naturally the missing baryon problem at low-$z$. We thus predict that the majority of the low-$z$ (local) baryon matter is not bounded by DM halos.

The measured PVD with 2dF and galaxy luminosity density at low-$z$ are consistent with $\Omega_{{\rm M},z}\sim 0.2$ at $z\lesssim 0.1$ \cite{Tinker2007,Keenan2012}. Different compilations of SNe~Ia data, joint fits between CMB data ({\textit{Planck}} or {\textit{WMAP}}) with other observations (including lensing and BAO) all support a picture of $\Omega_{{\rm M},z}$ increasing with $z$. This also explains why the {\textit{Planck}} results are consistent BAO data but not with SNe~Ia data, due to the fact the BAO measures mostly $\Omega_{{\rm M},z_2}$ and $h_{0,z_2}$ ($z_2\sim 1100$), though observed at very low-$z$. (Please refer to SM for extensive discussions.) Many cosmological probes, such as PVD, galaxy counts and luminosity function, clusters of galaxies, lensing, etc, can only measure DM content in cosmic structures. On the other hand, standard candles (e.g. SNe~Ia) and standard rulers (e.g. BAO) measure effectively the average DM content in the universe. Therefore $\Omega_{{\rm M},0}=0.030\pm 0.011$ at $z\ll 1$, inferred from PVD of SNe~Ia hosts, might indicate that the local DM is mostly not contained in DM halos, but distributed uniformly as a background of the local universe, in a similar way to the missed local baryons from cosmic structures. A significant amount of uniformly distributed DM at low-$z$ is, however, not expected at all in cold or even warm-DM models. Unfortunately, there is currently no effective way to measure the diffuse matter density in the local universe.

\acknowledgments{SNZ is grateful to Drs. Hong Li and Yu-Zhong Wu who helped me in collecting the data used in this work. Xuelei Chen, Yipeng Jing, Tipei Li, Weipeng Lin, Richard Lieu, Roger Penrose, David Valls-Gabaud, Junqing Xia, Marcel Zemp, Bing Zhang and Pengjie Zhang are thanked for discussions, comments or suggestions. SNZ acknowledges partial funding support by 973 Program of
China under grant 2009CB824800, by the National Natural Science Foundation of China under grant Nos. 11133002
and 10725313, and by the Qianren start-up grant 292012312D1117210.}

\newpage

\setcounter{figure}{0}

\begin{center}
{\large{\bf Supplementary material}}
\end{center}

\vspace{4mm}

\noindent {\bf 1. Data and sky map of the eight SNe Ia}
\\

\begin{figure}[bht]
\begin{center}
\includegraphics[angle=0,width=0.4\textwidth]{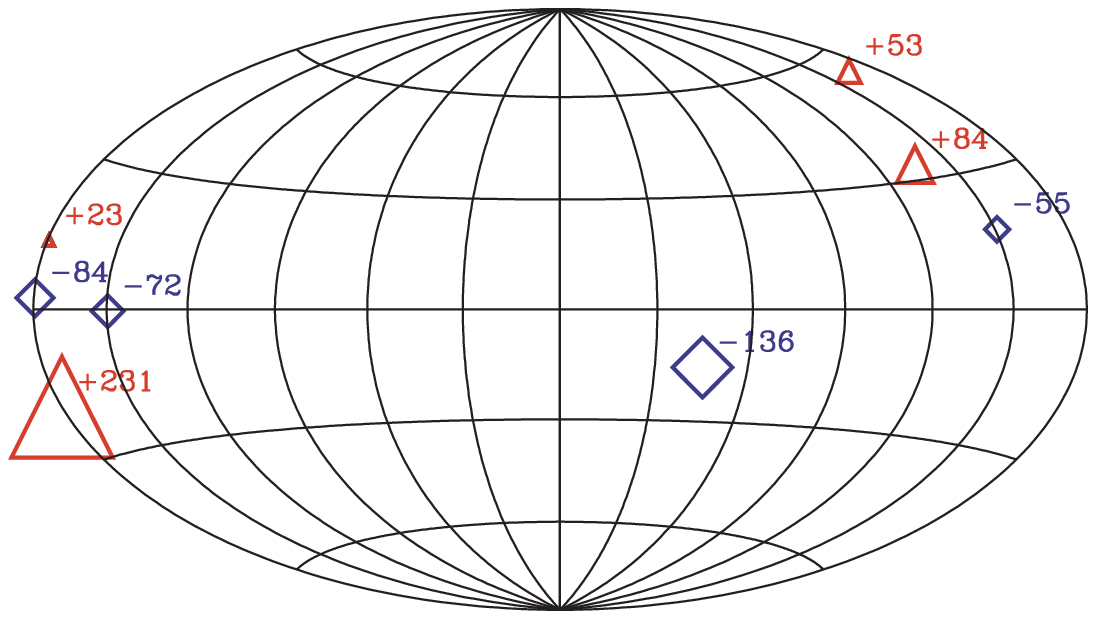}
\end{center}
\caption{Positions of the eight SNe~Ia in equatorial coordinates and their LOS peculiar velocities from the Hubble flow. Negative (marked as blue diamonds) and positive (marked as red triangles) velocities mean that their hosts are moving towards and away from the observer within the Hubble flow with $h_{0,0}=0.738$, respectively. The sizes of the signs are proportional to their LOS velocity deviations.} \label{fig:3}
\end{figure}

Table 1 lists all data of the eight SNe Ia used in this work. Figure~\ref{fig:3} plots in the sky map the positions and LOS velocities of the eight SNe Ia, which are consistent with random and isotropic distribution (albeit with small number statistics).
\begin{table*}
\caption{Redshift, coordinates (J2000.0), Hubble constant and LOS peculiar velocities of the eight SNe~Ia.}
\begin{center}
\begin{tabular}{lccccc}
\\
\hline \hline

Name & redshift & Right ascension & Declination & Hubble Constant& LOS pec. vel.\\
     &          & (hh:mm:ss)      & (dd:mm:ss)& $H_0$~(km~s$^{-1}$~Mpc$^{-1}$) & $V_{\rm los}$~(km~s$^{-1}$)\\
\hline \hline
 1981b  & 0.0072 & 12:34:29.57 &  +02:11:59.3   & 70.37 (2.26) &-84.7 (66.7)\\
 1990n  & 0.0043 & 12:42:56.68 &  +13:15:23.4   & 74.81 (2.75) &  23.3 (48.3)\\
 1994ae & 0.0045 & 11:56:25.87 &  +55:07:43.2   & 76.38 (2.46) &  53.2 (45.1)\\
 1998aq & 0.0055 & 10:47:01.94 &  +17:16:30.8   & 70.89 (2.61) & -55.9 (58.6)\\
 1995al & 0.0059 & 09:50:55.97 &  +33:33:09.4   & 76.87 (2.83) &  84.1 (68.1)\\
 2002fk & 0.0070 & 03:22:05.71 &  $-$15:24:03.2 & 68.48 (2.83) &-136.3 (81.1)\\
 2007af & 0.0070 & 12:01:52.80 &  $-$18:58:21.7 & 81.43 (2.62) & 231.3 (75.0)\\
 2007sr & 0.0063 & 14:22:21.03 &  $-$00:23:37.6 & 70.50 (2.59) & -72.5 (66.8)\\
\hline \hline
\end{tabular}
\end{center}
\end{table*}

\vspace{5mm}
\noindent {\bf 2. Pairwise velocity dispersion}
\\

\begin{figure}[bht]
\begin{center}
\includegraphics[angle=0,width=0.35\textwidth]{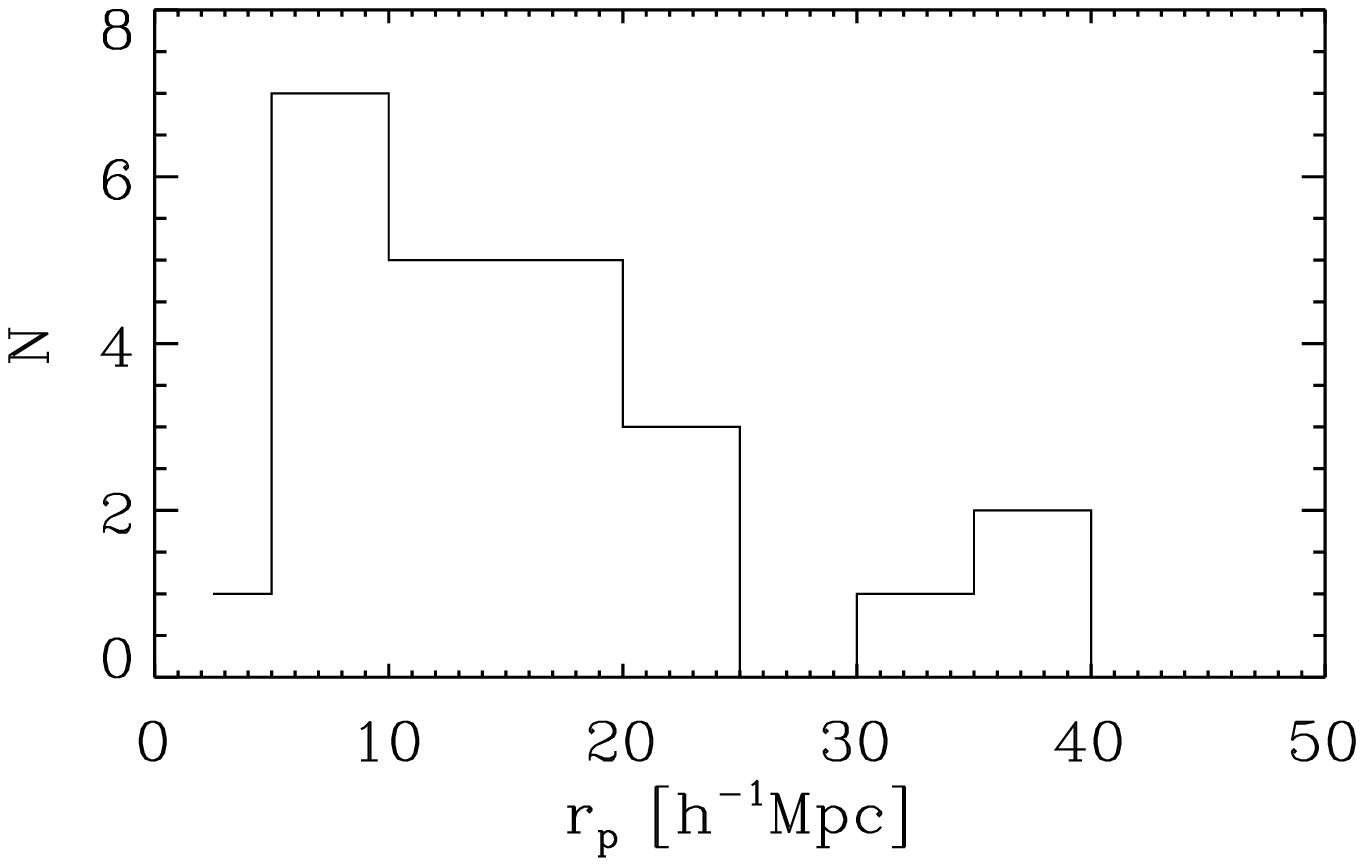}
\end{center}
\caption{Distribution of projected separations between the 28 pairs.} \label{fig:5}
\end{figure}

The apparent pairwise velocity dispersion is calculated between the eight SNe~Ia as
\begin{equation}\label{pairwise}
\sigma_{12}^2=\sum_{i,j}(V_{{\rm los},i}-V_{{\rm los},j})^2/N,
\end{equation}
where $i=1$ to 7, $j=i+1$ to 8, and $N=28$ is the total number of pairs. The distribution of the projected separations of these pairs is shown in Figure~\ref{fig:5}. The true pairwise velocity dispersion should be given by
\begin{equation}\label{true_pairwise}
\hat{\sigma}_{12}^2=\sigma_{12}^2-2\sigma_0^2,
\end{equation}
where $\sigma_0=63.8{\rm \ km \ s}^{-1}$ is the average measurement error in $V_{{\rm los},i}$.

\begin{figure}[bht]
\begin{center}
\includegraphics[angle=0,width=0.35\textwidth]{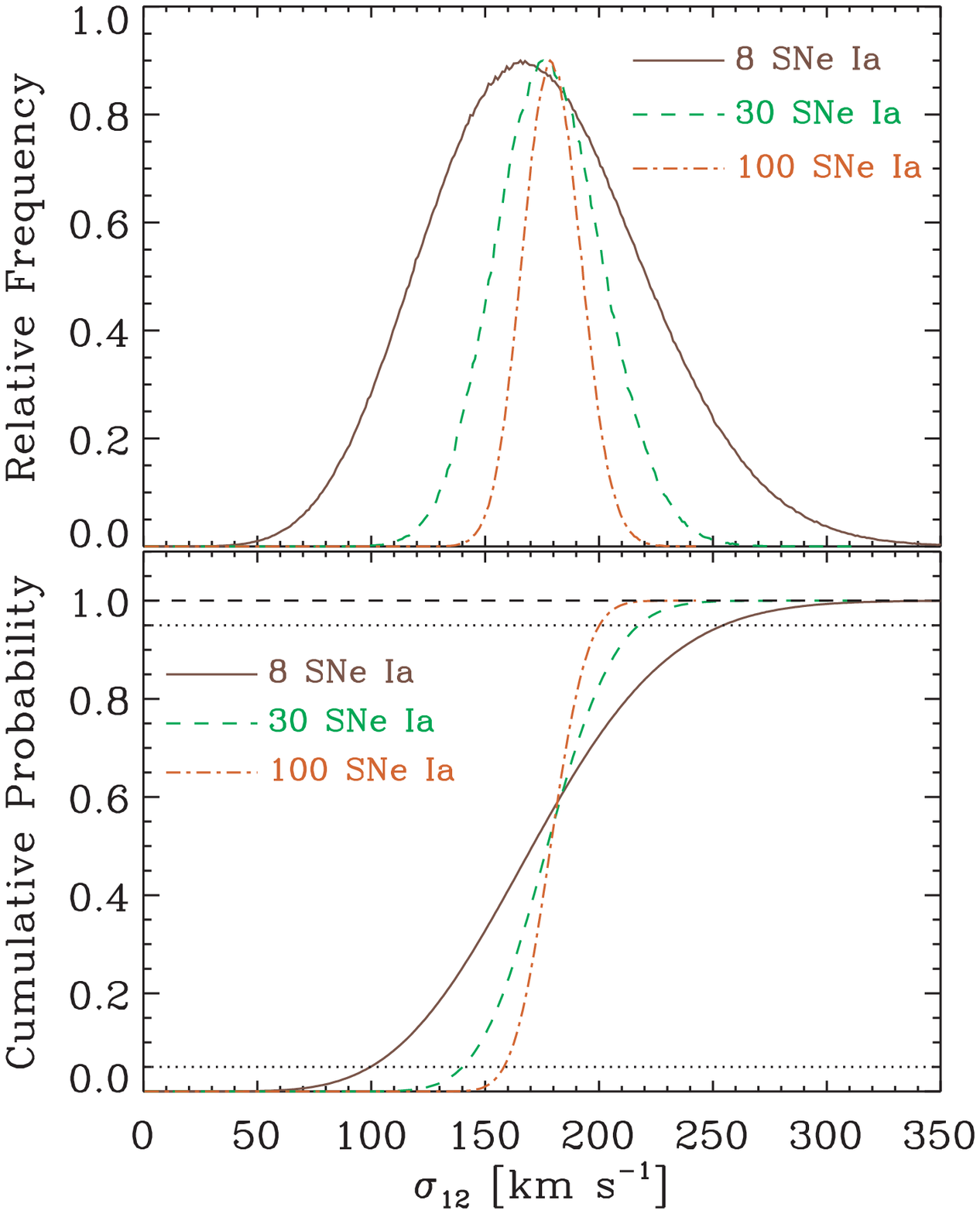}
\end{center}
\caption{Simulated distribution $\sigma_{12}$ and its cumulative probability distribution, for the cases of 8, 30 and 100 SNe Ia, respectively.} \label{fig:6}
\end{figure}

For this small sample of peculiar velocities with measurement errors, we perform Monte-Carlo simulations to evaluate the expected values of the pairwise velocity dispersion and its error. A random sample of $V_{{\rm los},i}$ is produced with a Gaussian distribution of zero mean and standard deviation $\sigma$. Then for each random group of eight $V_{{\rm los},i}$, $\sigma_{12}$ is calculated with equation~(\ref{pairwise}). The distribution of $\sigma_{12}$ and its cumulative probability distribution are shown in Figure~\ref{fig:6}; $\sigma=122.5{\rm \ km \ s}^{-1}$ is chosen so that the observed $\sigma_{12}=167.2{\rm \ km \ s}^{-1}$ equals the mean value of the simulated distribution. With equation~(\ref{true_pairwise}), we have $\hat{\sigma}_{12}=141\pm 26.5{\rm \ km \ s}^{-1}$ ($1\sigma$ range). For comparison, we also show the simulation results for 30 and 100 SNe Ia, which produce $\sigma(\hat{\sigma}_{12})=13$ and $\sigma(\hat{\sigma}_{12})=6$, respectively.

\vspace{5mm}
\noindent {\bf 3. Input parameters with errors and derived dark energy parameters}
\\

In Table~2, we list the input parameters for $h_{0,z_i}$ and $\Omega_{{\rm M},z_i}$ with their $1\sigma$ errors and the derived the dark energy parameters with 95\% confidence regions. The distributions of the derived the dark energy parameters shown in Figure~4 of the main paper are obtained with Monte-Carlo simulations as follows:
\begin{itemize}
\item We assume that the input parameters follow Gaussian distributions;
\item We sample the input parameters with the assumed Gaussian distributions one million times;
\item For each group of the sampled input parameters, we calculate $\Psi_{\Lambda,z}=(1-\Omega_{{\rm M},z})h_{0,z}^2$ and $3(1+w_{z_i,z_j})=-\log(\Psi_{\Lambda,z_i}/\Psi_{\Lambda,z_j})/\log((1+z_j)/(1+z_i))$;
\item We finally obtain the distributions of $\Psi_{\Lambda,z}$ and $3(1+w_{z_i,z_j})$ and their 95\% confidence region.
\end{itemize}

\begin{table*} \label{para}
\caption{Input parameters with errors and derived dark energy parameters.}
\begin{center}
\begin{tabular}{cccccc}
\hline \hline
\multicolumn{2}{c}{$z_0\sim 0$} & \multicolumn{2}{c}{$z_1\sim 1$} &\multicolumn{2}{c}{$z_2\sim 1100$} \\
\hline
$10^2\cdot h_{0,0}$ & $10^2\cdot \Omega_{{\rm M}, 0}$ &  $10^2\cdot h_{0,z_1}$ & $10^2\cdot \Omega_{{\rm M}, z_1}$ &  $10^2\cdot h_{0,z_2}$ & $10^2\cdot \Omega_{{\rm M}, z_2}h_{0,z_2}^2$ \\
$70.3\pm3.52$ & $3 \pm 1.1$ & $70.4\pm 0.51$ & $28 \pm 1$ & $67.9\pm 1.5$ & $14.23\pm0.29$\\
\hline
 \multicolumn{2}{c}{$\Psi_{\Lambda,0}=0.53_{-0.10}^{+0.21}$} &
 \multicolumn{2}{c}{$\Psi_{\Lambda,z_1}=0.357_{-0.014}^{+0.029}$}& \multicolumn{2}{c}{$\Psi_{\Lambda,z_2}=0.103_{-0.040}^{+0.082}$}\\
\hline
 \multicolumn{2}{c}{$-(w_{z_0,z_1}+1)=0.188_{-0.102}^{+0.194}$} &
 \multicolumn{2}{c}{$-(w_{z_1,z_2}+1)=0.0060_{-0.0068}^{+0.014}$}& \multicolumn{2}{c}{$-(w_{z_0,z_2}+1)=0.024_{-0.011}^{+0.022}$}\\
\hline
 \multicolumn{2}{c}{$P(w_{z_0,z_1}\ge -1)=1.5\times 10^{-4} (3.6\sigma)$} &
 \multicolumn{2}{c}{$P(w_{z_1,z_2}\ge -1)=4\times 10^{-2} (1.7\sigma)$}& \multicolumn{2}{c}{$P(w_{z_0,z_2}\ge -1)=10^{-5} (4.3\sigma)$}\\
\hline \hline
\end{tabular}
\end{center}
Note: Errors for input parameters are all $1\sigma$. Errors for $\Psi_{\Lambda,z_i}$ and $w_{z_i,z_j}$ are for 95\% confident level.
\end{table*}

\vspace{5mm}
\noindent {\bf 4. Impacts of the error in $h_0$ and sample size}
\\

\begin{figure}[bht]
\begin{center}
\includegraphics[angle=0,width=0.45\textwidth]{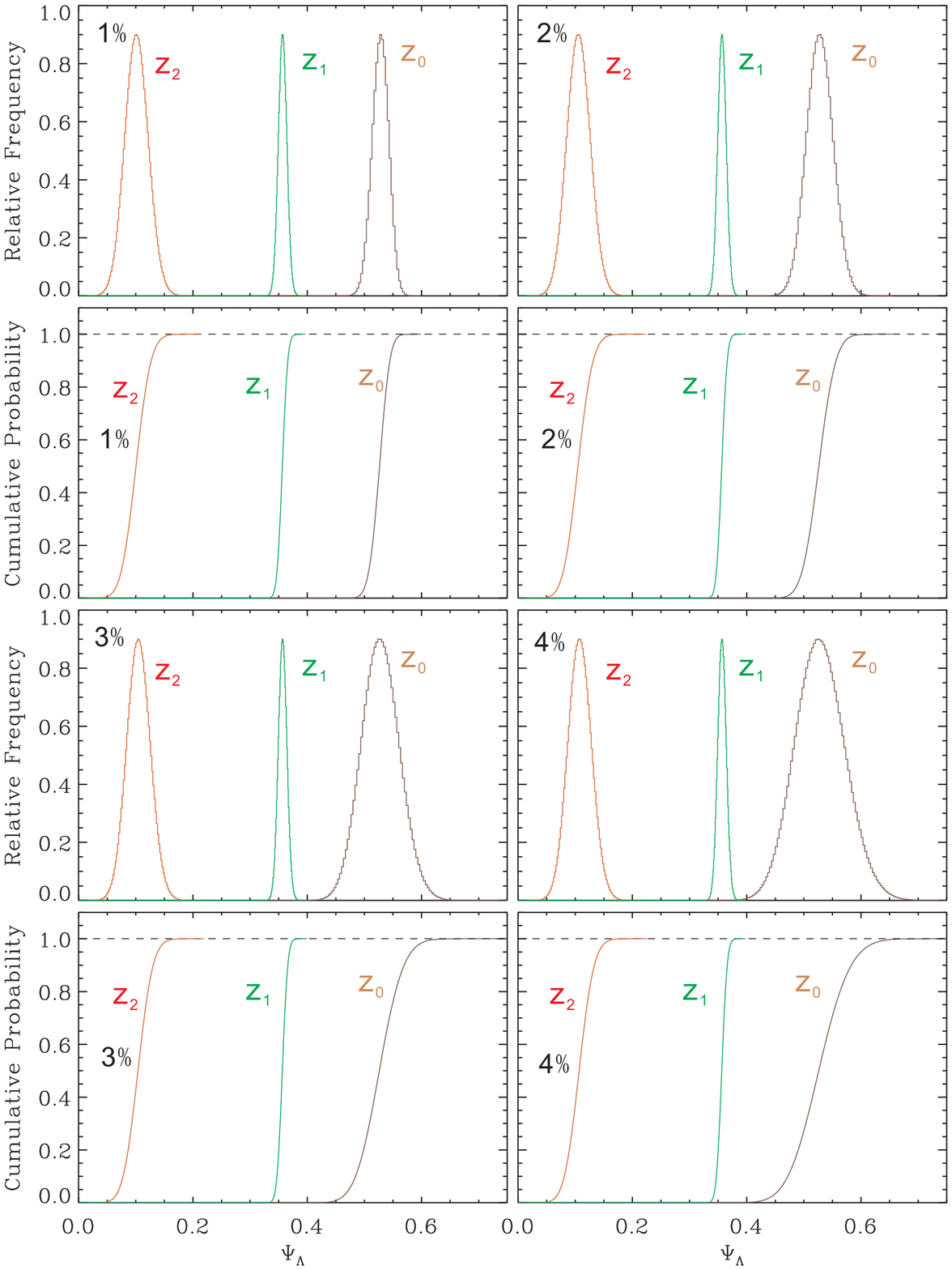}
\end{center}
\caption{The normalized dark energy density $\Psi_{\Lambda}$ obtained from measurements at three redshifts ($z_0\sim 0$, $z_1\sim 1$, and $z_2\sim 1100$) for different precisions of $h_{0,0}$. All other input parameters and errors are unchanged. Note that only $\Psi_{\Lambda,0}$ is changed.} \label{psi_all}
\end{figure}

\begin{figure}[bht]
\begin{center}
\includegraphics[angle=0,width=0.45\textwidth]{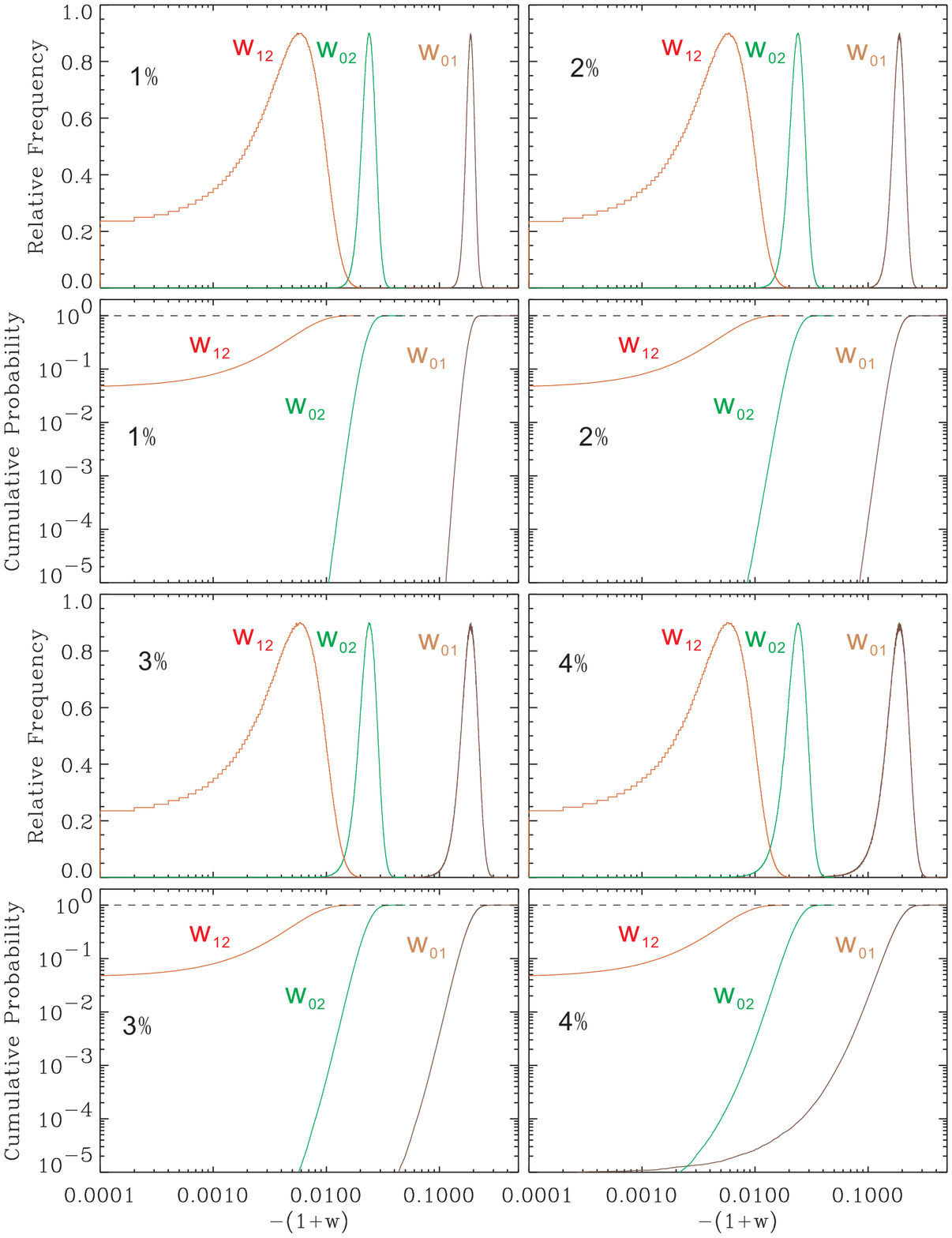}
\end{center}
\caption{The dark energy equation-of-state parameter $w$ obtained from measurements at three redshifts ($z_0\sim 0$, $z_1\sim 1$, and $z_2\sim 1100$) for different precisions of $h_{0,0}$. All other input parameters and errors are unchanged. Note that $w_{1,2}$ is unchanged.} \label{eos_all}
\end{figure}

In Figure~4 of the main paper, we conservatively assumed a 5\% error in $h_0$, which is reasonable due to the reported 3.3\% uncertainty \cite{supRiess2011} and a possible 2.5\% uncertainty due to cosmic variance \cite{supMarra}. To understand the impacts of the error in $h_0$ to the inferred dark energy parameters, in Figures~\ref{psi_all} and \ref{eos_all}, we show the derived distributions of normalized dark energy density $\Psi_\Lambda$ and equation-of-state parameter $w$. We can see dramatic improvements even if the error in $h_0$ is reduced by an additional 1\%.

For completeness, we also simulated the cases for samples of 30 and 100 SNe Ia, in order to overcome the possible error in $\hat{\sigma}_{12}$ caused by the small sample size of only eight SNe Ia in the current study. However, only very marginal improvements are expected even if the sample size is increased to 100, although the uncertainties to $\hat{\sigma}_{12}$ are significantly reduced, in proportion to $1/\sqrt{n}$, where $n$ is the sample size. This is because $\Omega_{{\rm M},0}$ is so small, that the dominant factor in determining $\Psi_{\Lambda,0}=(1-\Omega_{{\rm M},0})h_{0,0}^2$ is $h_{0,0}$. However, a larger sample will improve the statistical accuracy in $h_{0,0}$, again in proportion to $1/\sqrt{n}$, which will results in better determination of $\Psi_{\Lambda,0}$.

\vspace{5mm}
\noindent {\bf 5. Discussions on BAO measurements}

\vspace{5mm}
\noindent {5.1 Simplified case}
\\

The Baryon Acoustic Oscillation (BAO) observations measure the acoustic sound horizon size $d_{\rm H}$ at the last scattering surface ($z_{\rm L}=z_2 \sim 1100$), where the Cosmic Microwave Background (CMB) radiation is produced. With the base $\Lambda$CDM model and assuming that the cosmological parameters are the same from infinite redshift to redshift to $z_{\rm L}$, we have \cite{supWeinberg2008}
\begin{multline}\label{d_H}
d_{\rm H}=\frac{2}{H_{0,z_{\rm L}}\sqrt{3R_{\rm L}\Omega_{{\rm M},z_{\rm L}}(1+z_{\rm L})^3}} \\
\ln(\frac{\sqrt{1+R_{\rm L}}+\sqrt{R_{\rm EQ}+R_{\rm L}}}{1+\sqrt{R_{\rm EQ}}}),
\end{multline}
where $R\equiv 3\rho_{\rm B}/4\sigma T^4$, $\rho_{\rm B}$ is baryon density, and the subscripts `L' and `EQ' refer to the last scattering surface and matter-radiation equilibrium, respectively. Since $R_{\rm L}\propto \Omega_{\rm B}H_{0,z_{\rm L}}^2$, we have (aside from a slowly varying logarithm) \cite{supWeinberg2008}
\begin{equation}\label{d_H_sim}
d_{\rm H} \propto \Omega_{\rm B}^{-1/2}\Omega_{{\rm M},z_{\rm L}}^{-1/2}h _{0,z_{\rm L}}^{-2}.
\end{equation}
Therefore what BAO observations really measure are the cosmological parameters projected to $z=0$ from the last scattering surface, i.e., the same as that with CMB observations.

However, in order to obtain the BAO scale at a certain redshift $z$, the angular diameter distance at $z$ must be calculated, which depends on the cosmological parameters from zero redshift to $z$. For simplicity, we assume that the cosmological parameters (again in the base $\Lambda$CDM model) are the same from zero redshift to $z$. The angular diameter distance is then given by,
\begin{multline}
D_{\rm A}=\frac{c}{H_{0,z}(1+z)}\\
\int_0^z\frac{dz'}{\sqrt{\Omega_{{\rm M},z}(1+z')^3+\Omega_{K,z}(1+z')^2+\Omega_{\Lambda,z}}}.
\end{multline}

Therefore BAO data are connected to cosmological parameters both at redshift from zero to $z$ and at redshift from $z_{\rm L}$ to infinity. With BAO measurements alone, it is in principle not possible to obtain cosmological parameters at both ends, due to the degeneracy discussed above. However, combining with other low redshift data (such as SNe~Ia data discussed in this work), it is easy to break the degeneracy and to obtain cosmological parameters at higher than $z_{\rm L}$, which can be then compared with CMB results.

\begin{figure}[bht]
\begin{center}
\includegraphics[angle=0,width=0.4\textwidth]{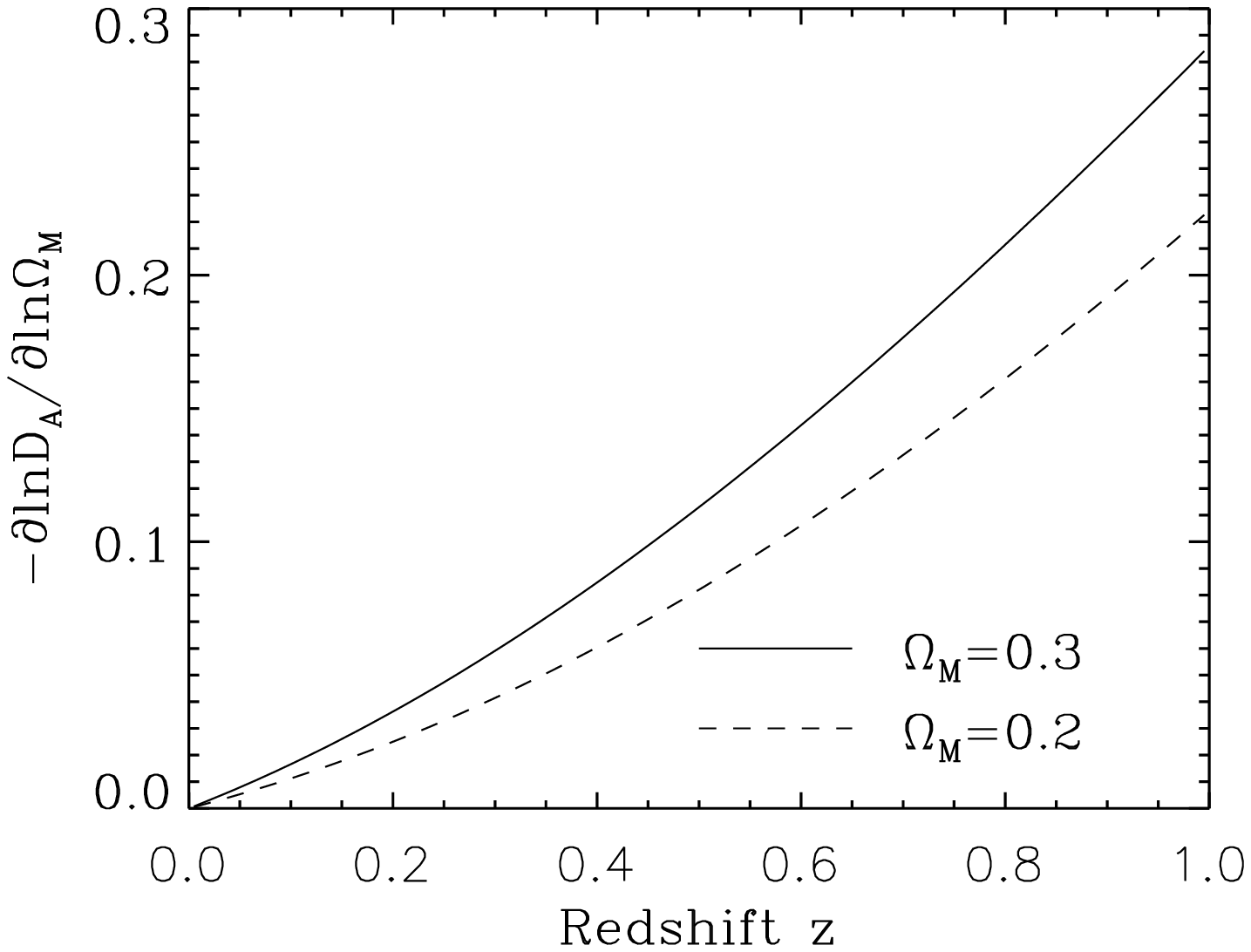}
\end{center}
\caption{Dependence of angular diameter distance on $\Omega_{{\rm M}}$ at different redshift; $\Omega_{K}=0$ is assumed.} \label{bao}
\end{figure}

It is easy to find: $-\partial \ln d_{\rm H}/\partial \ln h_{{\rm 0},z_{\rm L}}$=2, $-\partial \ln D_{\rm A}/\partial \ln h_{{\rm 0},z}$=1, $-\partial \ln d_{\rm H}/\partial \ln\Omega_{{\rm M},z_{\rm L}}$=0.5; $-\partial \ln D_{\rm A}/\partial \ln\Omega_{{\rm M},z}<0.3$ for $z<1$, as shown in Figure~\ref{bao} for $\Omega_{{\rm M},z}=0.2$ and 0.3 ($\Omega_K=0$ is assumed), respectively. This demonstrates that the BAO observations are always more sensitive to $h_{{\rm 0},z_{\rm L}}$ than to $h_{{\rm 0},z}$, and are more sensitive to $\Omega_{{\rm M},z_{\rm L}}$ than to $\Omega_{{\rm M},z_{\rm L}}$ when $z\lesssim1$.

\vspace{5mm}
\noindent {5.2 Realistic case}
\\

\begin{figure}[bht]
\centerline{\includegraphics[angle=0,width=0.4\textwidth]{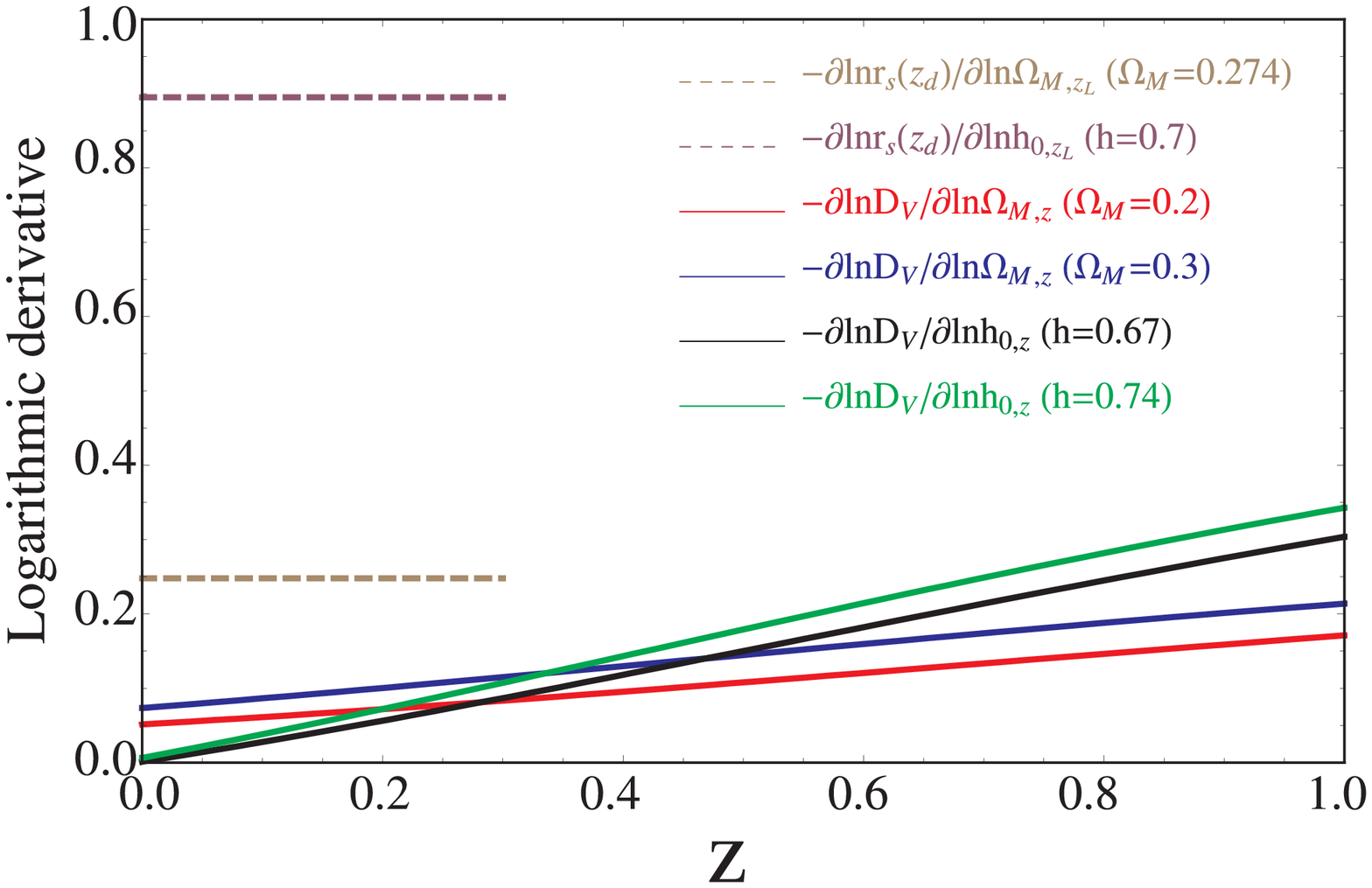}}
\caption{Dependence of $r_{\rm s}$ and
$D_{\rm V}(z)$ on $\Omega_{{\rm M}}$ and $h_0$; $\Omega_{K}=0$ is assumed.}
\label{bao1}
\end{figure}

Realistically, we need to consider the logarithm term in equation~(\ref{d_H}), the difference between the drag epoch ($z_{\rm d}$) and the epoch at the last scattering surface ($z_{\rm L}$), and the fact that BAO measurements are always obtained within a certain spherical volume of redshift $z$, which can be taken into account by the spherically averaged $D_{\rm A}$, i.e., $D_{\rm V}$. Therefore what BAO experiments actually measure is $r_{\rm s}(z_{\rm d})/D_{\rm V}(z)$ (Ref.\cite{supHinshaw2012}). We use
Eqs.~(1)-(6) in Ref.\cite{supKomatsu2008} to calculate $r_{\rm s}$ and
$D_{\rm V}(z)$. Similar to the simplified case, we again take logarithmic derivative to parameters
$\Omega_{\rm M}$ and $h_0$; the results are
shown in Figure~\ref{bao1}.

One can see that $-\partial \ln r_{\rm s}(z_{\rm d})/\partial \ln \Omega_{{\rm M},z_{\rm L}}=0.248$ and $-\partial \ln r_{s}(z_{d})/\partial \ln
h_{0,z_{\rm L}}=0.895$, consistent to the previously used (e.g., Eq.~(12) of Ref.\cite{supPercival2009})
 \begin{equation}\label{d_H_rea}
 r_{\rm s}(z_{\rm d})\propto \Omega_{{\rm M}}^{-0.255}h_0^{-0.778},
\end{equation}
 which is quite different from Eq.~(\ref{d_H_sim}) for the simplified case discussed above. Nevertheless, similarly to the simplified case, our details calculation also shows $-\frac{\partial \ln r_{\rm s}(z_{\rm d})}{\partial \ln \Omega_{{\rm M},z_{\rm L}}}> -\frac{\partial \ln D_{\rm V}(z_1)}{\partial \ln
\Omega_{{\rm M},z}}$ and $-\frac{\partial \ln r_{\rm s}(z_{\rm d})}{\partial \ln h_{0,z_{\rm L}}}> -\frac{\partial \ln D_{\rm V}(z)}{\partial \ln h_{0,z}}$ ($z < 1$); therefore connecting $D_{\rm V}(z)$ directly to $r_{\rm s}(z_{\rm d})$, i.e., assuming $D_{\rm V}(z)\propto r_{\rm s}(z_{\rm d})$, actually probes $\Omega_{{\rm M},z_{\rm L}}$ and $h_{0,z_{\rm L}}$ much more sensitively than to $\Omega_{{\rm M},z}$ and $h_{0,z}$ ($z<1$), unless $\Omega_{{\rm M},z_{\rm L}}=\Omega_{{\rm M},z}$ and $h_{0,z_{\rm L}}=h_{0,z}$, as assumed implicitly in previous studies.

Since all existing BAO observations are made to $z<1$, the above analyses explain naturally why $h_0$ and $\Omega_{{\rm M}}$ obtained with BAO data are consistent with CMB results, but not with SNe~Ia results with data at even similar redshifts. Nevertheless BAO data are connected to cosmological parameters ($\Omega_{{\rm M}}$ and $h_0$ here) at both $z_{\rm L}$ and at low-$z$. In fact, the combined-BAO derived $H_0=68.4^{+1.0}_{1.0}$ and $\Omega_{{\rm M}}=0.305^{+0.009}_{0.008}$ (the last row in Table 8 of Ref.\cite{supBielewicz2013}) are between that of SNe Ia and {\textit{Planck}}-only results but are closer to the latter, in support to our above analysis.

\vspace{5mm}
\noindent {\bf 6. Low redshift matter density}
\\

In the paper, we have assumed the matter density $\Omega_{{\rm M},z_1}=0.28\pm 0.01$, which is consistent with the high-$z$ ($z>0.5$) Union 2.1 SNe Ia data and has been well-accepted in the concordance $\Lambda$CDM model prior to the {\textit{Planck}} result. Both CMB and BAO data contribute significantly to the this value, since it has been assumed so far that globally there is a unique $\Omega_{{\rm M}}$. However, neither CMB nor BAO data should be used to measure uniquely $\Omega_{{\rm M},z_1}$, as discussed above.

There are still many other cosmological probes that can be used to measure $\Omega_{{\rm M},z_1}$, such as galaxy counts, galaxy luminosity functions, weak gravitational lensing, peculiar velocities of galaxies inferred from redshift space distortion with two-point correlation function, standard candles or rulers, etc. However galaxy counts and galaxy luminosity functions are known to be biased probes of matter density, due to redshift-dependent galaxy formation processes. On the other hand, weak gravitational lensing and peculiar velocities probe directly the gravitational field, and are thus neither biased nor redshift-dependent, when used to probe the matter density. Standard candles (such as SNE Ia) or rulers (such as BAO) have been extensively discussed above.
\\

\vspace{5mm}
\noindent {6.1 Joint fits between CMB and other data}
\\

It is instructive to inspect how the best fit of $\Omega_{{\rm M}}$ changes, by combining CMB temperature power-spectrum with other probes. In Table 3, we compile such a list from {\textit{Planck}} results \cite{supBielewicz2013} for easy comparison. From this table, we notice the following:
\begin{enumerate}
\item For the first two groups on the {\textit{Planck}} data, i) all combinations results in lower $\Omega_{{\rm M}}$, higher
$h_0$ and $\Psi_\Lambda$, consistent with our model that the dark energy density increases with cosmic time; ii) the percentages of changes are larger when combined with lensing, SNLS, or HST data; this is understandable since these data are only connected to low-$z$ cosmological parameters; iii) the combination with Union2 data has the smallest change, even smaller than that with BAO data. The last one can also be understood, because there are 29 SNe Ia with $z>1$ in the Union2 data, in comparison to only 11 SNe Ia with $z>1$ in the SNLS data and for all BAO data $z<1$. Therefore Union2 data probe $\Omega_{{\rm M}}$ at higher redshift than all other data, except the CMB data. Consequently these results suggest that $\Omega_{{\rm M},z}$ decreases with decreasing $z$ (or increasing cosmic time), once again fully consistent with our scenario.
\item For the last group on the {\textit{WMAP}} data, i) aside from the BAO combination, the signs of changes are the same as the first two groups, but the fractions of changes are far more significant, which can be understood since the signal-to-noise ratios of the {\textit{WMAP}} data are far less than that of the {\textit{Planck}} data, such that the {\textit{WMAP}} data play less significant roles in determining the joint fitting results; ii) for the BAO combination, the signs of changes are opposite to that on the {\textit{Planck}} data, inconsistent with our analysis that BAO data should find slightly lower $\Omega_{{\rm M}}$ and higher $h_0$. Therefore we suggest that the discrepancy between {\textit{WMAP}} and {\textit{Planck}} data on  $\Omega_{{\rm M}}$ and $h_0$ may be due to systematic errors in {\textit{WMAP}} data analysis. As a matter of fact, previous independent re-analysis  \cite{supLiu2009} of {\textit{WMAP}} data have found consistent results with that released recently by the {\textit{Planck}} team.
\end{enumerate}

\begin{table} \label{planck}
\begin{small}
\caption{Matter density and Hubble constant obtained with different combinations of data in the base $\Lambda$CDM model \cite{supBielewicz2013}. The first row of each group lists values of the best-fit parameters, below which are the relative changes of these parameters in percentage when combined with other data. The last column is the normalized dark energy density defined in this paper: $\Psi_\Lambda\equiv \Omega_\Lambda h_{0}^2=(1-\Omega_{{\rm M}})h_{0}^2$ (for $\Omega_K=0$).}
\begin{center}
\begin{tabular}{lcccc}
\hline \hline
Data &  $\Omega_{{\rm M}}$ & $H_{0}$ & $\Omega_{{\rm M}}h_{0}^2$ & $\Psi_\Lambda$\\
\hline
lowL+lowLike &    0.318300   &   67.0400   &  0.143050  &       0.306386   \\
+ lensing (\%)  &    -4.71254   &   1.67065  &   -1.49598   &           5.64073    \\
+ BAO (\%)  &     -2.51336  &   0.880066  &  -0.789933  &               2.96209    \\
+ HST (\%)  &     -5.46654  &    1.96897  &   -1.71269  &               6.63307    \\
+ SNLS (\%)  &       -3.58153   &   1.25298   &  -1.15345  &            4.23755    \\
+ Union2 (\%)  &       -1.53943  &   0.551913  &  -0.433420  &          1.82602    \\
\hline
lowL+lowLike+highL   &   0.317000   &   67.1500  &   0.142970  & 0.307942   \\
+ lensing (\%)  &       -3.43848  &    1.17647  &   -1.16108  &         4.00467    \\
+ BAO (\%)  &      -2.64984   &  0.923299 &   -0.888296   &             3.12879    \\
+ HST (\%)  &       -4.76341  &    1.66790  &   -1.58075  &             5.65917    \\
+ SNLS (\%)  &       -3.47003   &   1.20625   &  -1.15408 &             4.08967    \\
+ Union2 (\%)  &       -1.45110  &   0.506324  &  -0.489611  &          1.71384    \\
\hline
{\textit{WMAP}}   &   0.292000  &    68.8700   &  0.138400  &                      0.335908   \\
+ BAO (\%)  &        1.84932  &  -0.580807  &   0.794800  &             -1.96292   \\
+ HST (\%)  &      -10.0685    &  3.81878   &  -2.96243  &              12.2109    \\
+ SNLS (\%)  &       -15.1712   &   4.82068   &  -5.46965  &            15.8555    \\
+ Union2 (\%)  &       -11.6096  &    3.57195   &  -4.16907  &          11.7333    \\
\hline \hline
\end{tabular}
\end{center}
Notes: {\bf lowL}: low-$l$ {\textit{Planck}} temperature ($2\leq l \leq 49$); {\bf highL}: high-$l$ {\textit{Planck}} temperature (CamSpec, $50\leq l \leq 2500$); {\bf lensing}: {\textit{Planck}} lensing power spectrum reconstruction; {\bf lowLike}: low-$l$ {\textit{WMAP}} 9 polarization; {\bf BAO}: Baryon oscillation data from DR7, DR9 and and 6DF; {\bf SNLS}: Supernova data from the Supernova Legacy Survey;
{\bf Union2}: Supernova data from the Union compilation;
{\bf HST}: Hubble parameter constraint from HST (Riess et al. \cite{supRiess2011});
{\bf {\textit{WMAP}}}: The full {\textit{WMAP}} (temperature and polarization) 9 year data.
\end{small}
\end{table}

\vspace{15mm}
\noindent {6.2 SNe Ia data}
\\

In Figure 19 of Ref.\cite{supBielewicz2013}, the best fit $\Omega_{{\rm M}}$ is found to increase from SNLS, to Union2 and {\textit{Planck}} data, consistent with the fact that these data probe $\Omega_{{\rm M}}$ with increasing redshift. Therefore $\Omega_{{\rm M},z_1}=0.28$ ($z_1\sim 1$) is a reasonable choice, determined with the Union2 data. On the other hand, it is quite possible that the SNLS data probe $\Omega_{{\rm M}}$ at lower redshift more sensitively, with $\Omega_{{\rm M},z}=0.23$ at $z\sim 0.5$.

\vspace{5mm}
\noindent {6.3 Peculiar velocities from 2dF survey}
\\

In Figure 6 of Ref.\cite{supTinker2007}, the distribution of pairwise peculiar velocity dispersion (PVD) of galaxies in the 2dF survey is compared with high-resolution $N$-body simulation. The flattening of the observed PVD to about 450~km~s$^{-1}$ to the projected separations above several Mpc is consistent with $\Omega_{{\rm M}}=0.2$. Since the median redshift of the 2dF survey is 0.1, this result suggest that $\Omega_{{\rm M},z}=0.2$ at $z\sim 0.1$, again suggesting increasing $\Omega_{{\rm M}}$ with redshift.
\\

\vspace{5mm}
\noindent {\bf 7. Local matter density}
\\

In the paper, we have found a very low local matter density $\Omega_{{\rm M},0}=0.03\pm0.011$ at about 20~$h^{-1}$Mpc, i.e., just outside our local group. Here we examine critically if this is consistent with other observations.

Local galaxy counts within 5~$h^{-1}$Mpc show an over-density of about 25\%, compared to the cosmological average \cite{supSchlegel1994}. This may argue for an over-density in the total local mass distribution. However, this is not in conflict with our conclusion of low $\Omega_{{\rm M}}$ in present day universe, since the over-density is just outside the local group and well-within the local supercluster, where the mean matter density should be higher than the cosmological average. The total baryon mass density derived from the most complete nearby galaxy catalogue is $\Omega_{{\rm B}}=0.023$ within about 6~$h^{-1}$Mpc, about half of the cosmological average \cite{supKarachentsev2004}. Since the mean HI density decreases rapidly at larger distances (Figure 14 of Ref. \cite{supKarachentsev2004}), it is physically plausible that $\Omega_{{\rm B}}$ decreases by a factor of a few at about 20~$h^{-1}$Mpc, i.e., nearly to the boundary of the local supercluster. Therefore $\Omega_{{\rm M},0}=0.03\pm0.011$ outside the local supercluster does not conflict the data of nearby galaxy data, if baryon to total mass ratio remains about the same as the cosmological average. Actually the very low PVD of about 100~km~s$^{-1}$ of these galaxies within about 6~$h^{-1}$Mpc at projected separations of below and around 1~$h^{-1}$Mpc also supports a very low value of $\Omega_{{\rm M}}$.

In Figure 6 of Ref.\cite{supKeenan2012}, a significant lower luminosity density of galaxies is found at redshift between 0.02 to 0.07, compared to that above redshift of 0.1. Even after considering possible cosmic variance, a significant lower mass density at redshift below 0.1 cannot be excluded \cite{supKeenan2012}. This is also consistent with a significantly lower mass density in present day universe discussed above.

The data shown above, from high redshift to low redshift, all support a
picture that $\Omega_{{\rm M},z}$ increases with $z$. On the other hand, at low
redshift regime, baryonic matter traces the dark matter halos, so
$\Omega_{{\rm B},0}$ measured in cosmic structures should be also much lower than the global value. Since the total
baryon is conserved, this indicates that at low redshift and present day
Universe, most baryonic matter is not
bounded by dark matter halos and thus distributed
between galaxies as intergalactic medium. This scenario provides a natural explanation
of the missing baryon problem and can be tested with future observations.
\\

\vspace{5mm}
\noindent {\bf 8. $\Lambda$LTB$(t)$ scheme}
\\

Our main conclusion is that the dark energy density increases with cosmic time, in a way that its equation-of-state parameter decreases with cosmic time and is less than $-1$ at low redshift. However, currently there is no well-understood physics to account for the dark energy evolving this way.

If matter and dark energy are coupled, then phenomenologically the increasing dark energy density would require decreasing matter content in the universe. A class of models, referred to as unified or coupled dark matter and dark energy models, have been widely studied in literature, such as the generalized Chaplygin gas model \cite{supBento2002} or gravity-dark energy coupling model \cite{supLi2013}. Unfortunately both models generally require a rapid increase of $\rho_\Lambda(z)$ with $z$, opposite to our result. Our results that $\rho_\Lambda(z)$ increases with cosmic time seems to imply that dark matter is continuously converted to dark energy and the conversion rate only increases rapidly at low-$z$, which drives the observed super-accelerating expansion of the universe. Mathematically a cosmological model of universe with time varying parameters can be described with the LTB metric including the dark energy term $\Lambda$, the so-called $\Lambda$LTB metric. We name this class of models as $\Lambda$LTB$(t)$ models, to distinguish them from the $\Lambda$LTB models commonly used to describe a spherically symmetric, but spatially inhomogeneous universe, i.e, $\Lambda$LTB$(r)$ models. Here we suggest a specific $\Lambda$LTB$(t)$ scheme in which the time varying parameter is $\rho_\Lambda(z)$ (or $\Psi_{\Lambda,z}$). Within this framework, the $\Lambda$CDM model can be used to describe the universe at any epoch of cosmic time.

A generic property of a LTB$(t)$ model (with or without $\Lambda$) is that cosmological redshift $z$ is interpreted as the time coordinate $t$ (clock of the universe) rather than radial coordinate $r$. Therefore there is no such concept as radial inhomogeneity in a LTB$(t)$ model; any previously observed radial dependence of any cosmological parameter or physical quantities (even in the comoving frame) is interpreted as evolutionary effects in a LTB$(t)$ model. In principle simultaneous events can only be observed between two points in the universe with the same redshift (i.e. in the transverse direction), unless any evolutionary effect is negligible. The Copernican Principle is naturally maintained, since in a LTB$(t)$ model any observer anywhere in the universe observes at the same cosmic time the same thing in the universe. The Copernican Principle can be violated only if cosmological parameters or measured physical quantities show large scale anisotropy; in this case the spherical symmetry in the LTB prescription is broken. Therefore LTB$(t)$ models can be falsified by any observed large scale anisotropy.

\def\apj{Astrophys. J.}
\def\apjs{Astrophy. J. Suppl.}
\def\aj{Astron. J.}
\def\aap{Astron. Astrophys.}
\def\mnras{Mon. Not. R. Astron. Soc.}
\def\araa{Ann. Rev. Astron. Astrophys.}
\def\pasj{Pub. Astron. Soc. Jap.}


{\footnotesize
\begin{thebibliography}{10}

\bibitem{Riess2011}
A.~G. Riess, {\it et~al.\/}, {\it The Astrophysical Journal\/} {\bf 730}, 119
  (2011).

\bibitem{Riess2012}
A.~G. Riess, J.~Fliri, D.~Valls-Gabaud, {\it The Astrophysical Journal\/} {\bf
  745}, 156 (2012).

\bibitem{Riess1998}
A.~G. Riess, {\it et~al.\/}, {\it The Astronomical Journal\/} {\bf 116}, 1009
  (1998).

\bibitem{Perlmutter1999}
S.~Perlmutter, {\it et~al.\/}, {\it The Astrophysical Journal\/} {\bf 517}, 565
  (1999).

\bibitem{PlanckCollaboration2013}
{Planck Collaboration I}, P. A. R.~Ade, {\it et~al.\/}  (2013) (arXiv:1303.5062).

\bibitem{Suzuki2012}
N.~Suzuki, {\it et~al.\/}, {\it The Astrophysical Journal\/} {\bf 746}, 85
  (2012).

\bibitem{Zehavi1998}
I.~Zehavi, A.~G. Riess, R.~P. Kirshner, A.~Dekel, {\it The Astrophysical
  Journal\/} {\bf 503}, 483 (1998).

\bibitem{Alnes2006}
H. Alnes, M.~Amarzguioui, O. Gr\o~n, {\it Physical Review D\/} {\bf 73},
  083519 (2006).

\bibitem{Alexander2009}
S.~Alexander, T.~Biswas, A.~Notari, D.~Vaid, {\it Journal of Cosmology and
  Astroparticle Physics\/} {\bf 2009}, 025 (2009).

\bibitem{Mattsson2009}
T.~Mattsson, {\it General Relativity and Gravitation\/} {\bf 42}, 567 (2009).

\bibitem{Mortonson2009}
M.~Mortonson, W.~Hu, D.~Huterer, {\it Physical Review D\/} {\bf 80}, 067301
  (2009).

\bibitem{Sinclair2010}
B.~Sinclair, T.~M. Davis, T.~Haugb\o~lle, {\it The Astrophysical Journal\/}
  {\bf 718}, 1445 (2010).

\bibitem{Moss2011}
A.~Moss, J.~P. Zibin, D.~Scott, {\it Physical Review D\/} {\bf 83}, 103515
  (2011).

\bibitem{Ellis2011}
G.~F.~R. Ellis, {\it Classical and Quantum Gravity\/} {\bf 28}, 164001 (2011).

\bibitem{Nadathur2011}
S.~Nadathur, S.~Sarkar, {\it Physical Review D\/} {\bf 83}, 063506 (2011).

\bibitem{Marra2011}
V.~Marra, A.~Notari, {\it Classical and Quantum Gravity\/} {\bf 28}, 164004
  (2011).

\bibitem{Hinshaw2012}
G.~Hinshaw, {\it et~al.\/} (2012) (arXiv:1212.5226).

\bibitem{Riess2009}
A.~G. Riess, {\it et~al.\/}, {\it The Astrophysical Journal\/} {\bf 699}, 539
  (2009).

\bibitem{Hui2006}
L.~Hui, P.~Greene, {\it Physical Review D\/} {\bf 73}, 123526 (2006).

\bibitem{Zhang2008}
P.~Zhang, X.~Chen, {\it Physical Review D\/} {\bf 78}, 023006 (2008).

\bibitem{Tinker2007}
J.~L. Tinker, P.~Norberg, D.~H. Weinberg, M.~S. Warren, {\it The Astrophysical
  Journal\/} {\bf 659}, 877 (2007).

\bibitem{Marra}
V.~Marra, L.~Amendola, I.~Sawicki, W.~Valkenburg  (2013) (arXiv:1303.3121).

\bibitem{Zhang2011c}
P.~Zhang, A.~Stebbins, {\it Physical Review Letters\/} {\bf 107}, 041301
  (2011).

\bibitem{Zibin2011}
J.~P. Zibin, A.~Moss, {\it Classical and Quantum Gravity\/} {\bf 28}, 164005
  (2011).

\bibitem{Wang2013a}
F.~Y. Wang, Z.~G. Dai  (2013) (arXiv:1304.4399).

\bibitem{Schlegel1994}
D.~Schlegel, M.~Davis, F.~Summers, J.~A. Holtzman, {\it The Astrophysical
  Journal\/} {\bf 427}, 527 (1994).

\bibitem{Karachentsev2004}
I.~D. Karachentsev, V.~E. Karachentseva, W.~K. Huchtmeier, D.~I. Makarov, {\it
  The Astronomical Journal\/} {\bf 127}, 2031 (2004).

\bibitem{Keenan2012}
R.~C. Keenan, {\it et~al.\/}, {\it The Astrophysical Journal\/} {\bf 754}, 131
  (2012).

\end{thebibliography}
}


{\footnotesize
\begin{thebibliography}{10}
\bibitem{supRiess2011}
A.~G. Riess, {\it et~al.\/}, {\it The Astrophysical Journal\/} {\bf 730}, 119
  (2011).

\bibitem{supMarra}
V.~Marra, L.~Amendola, I.~Sawicki, W.~Valkenburg, {\it Physical Review Letters\/} {\bf 110}, 241305 (2013)

\bibitem{supWeinberg2008}
S.~Weinberg, {\it {Cosmology}\/} (Oxford University Press, Oxford, England,
  2008).

\bibitem{supHinshaw2012}
G.~Hinshaw, {\it et~al.\/} (2012) (arXiv:1212.5226).

\bibitem{supKomatsu2008}
E.~Komatsu, {\it et~al.\/}, {\it The Astrophysical Journal Supplement Series\/}
  {\bf 180}, 330 (2009).

\bibitem{supPercival2009}
W.~J. Percival, {\it et~al.\/}, {\it Monthly Notices of the Royal Astronomical
  Society\/} {\bf 401}, 2148 (2010).

\bibitem{supBielewicz2013}
{Planck Collaboration XVI}, P.~Bielewicz, {\it et~al.\/}  (2013) (arXiv:1303.5076).

\bibitem{supLiu2009}
H.~Liu, T.-P. Li (2009) (arXiv:0907.2731).

\bibitem{supTinker2007}
J.~L. Tinker, P.~Norberg, D.~H. Weinberg, M.~S. Warren, {\it The Astrophysical
  Journal\/} {\bf 659}, 877 (2007).

\bibitem{supSchlegel1994}
D.~Schlegel, M.~Davis, F.~Summers, J.~A. Holtzman, {\it The Astrophysical
  Journal\/} {\bf 427}, 527 (1994).

\bibitem{supKarachentsev2004}
I.~D. Karachentsev, V.~E. Karachentseva, W.~K. Huchtmeier, D.~I. Makarov, {\it
  The Astronomical Journal\/} {\bf 127}, 2031 (2004).

\bibitem{supKeenan2012}
R.~C. Keenan, {\it et~al.\/}, {\it The Astrophysical Journal\/} {\bf 754}, 131
  (2012).

\bibitem{supBento2002}
M.~Bento, O.~Bertolami, A.~Sen, {\it Physical Review D\/} {\bf 66}, 043507
  (2002).

\bibitem{supLi2013}
T.-P. Li, M.~Wu  (2013) (arXiv:0907.2731).

\end{thebibliography}
}
\end{document}